\documentclass[aps,prb,twocolumn,superscriptaddress,showpacs,floatfix]{revtex4-1}
\usepackage{epstopdf}
\usepackage{graphicx}
\usepackage{dcolumn}
\usepackage{bm}
\usepackage{bbm}

\usepackage{amsfonts}
\usepackage{amsmath}

\usepackage[colorlinks=true,citecolor=blue]{hyperref}
\hypersetup{colorlinks=true,citecolor=blue,linkcolor=red,urlcolor=blue}

\begin{document}
\renewcommand{\thefigure}{\arabic{figure}}
\def\be{\begin{equation}}
\def\ee{\end{equation}}
\def\ber{\begin{eqnarray}}
\def\eer{\end{eqnarray}}

\def\kv{{\bf k}}
\def\qv{{\bf q}}
\def\pv{{\bf p}}

\def\sigmav{{\bf \sigma}}
\def\tauv{{\bf \tau}}

\newcommand{\h}[1]{{\hat {#1}}}
\newcommand{\hdg}[1]{{\hat {#1}^\dagger}}
\newcommand{\bra}[1]{\left\langle{#1}\right|}
\newcommand{\ket}[1]{\left|{#1}\right\rangle}

\title{ Effective lattice Hamiltonian for monolayer MoS$_2$ : Tailoring electronic structure with perpendicular electric and magnetic fields}
\date{\today}

\author{Habib Rostami}
\affiliation{School of Physics, Institute for Research in Fundamental Sciences (IPM), Tehran 19395-5531, Iran}
\author{Ali G. Moghaddam}
\affiliation{Department of physics, Institute for Advanced Studies in Basic Sciences (IASBS), Zanjan 45137-66731, Iran}
\affiliation{School of Physics, Institute for Research in Fundamental Sciences (IPM), Tehran 19395-5531, Iran}
\author{Reza Asgari}
\email{asgari@ipm.ir} \affiliation{School of Physics, Institute
for Research in Fundamental Sciences (IPM), Tehran 19395-5531,
Iran}

\begin{abstract}
We propose an effective lattice Hamiltonian for monolayer MoS$_2$
in order to describe the low-energy band structure and investigate the effect of
perpendicular electric and magnetic fields on its electronic
structure. We derive a tight-binding model based on the
hybridization of the $d$ orbitals of molybdenum and $p$ orbitals of
sulfur atoms and then introduce a modified two-band continuum model of monolayer
MoS$_2$ by exploiting the quasi-degenerate partitioning
method. Our theory proves that the low-energy excitations of the system are no longer massive Dirac fermions. It reveals a difference between electron and hole
masses and provides trigonal warping effects.
Furthermore, we predict a valley degeneracy breaking effect in the
Landau levels. Besides, we also show that applying a gate voltage perpendicular
to the monolayer modifies
the electronic structure including the band gap and effective
masses.
\end{abstract}

\maketitle

\section{Introduction}
Although studies of two
dimensional (2D) electronic systems go back to some decades,
it was only in 2004 that the first truly 2D one-atom thick material, graphene, was isolated successfully~\cite{novoselov04}. Since
then the fundamental interest besides the promising applications in
nanoelectronic devices, has boosted the research about atomically
thin 2D materials. It has been recently demonstrated
that monolayer molybdenum misulfide (ML-MDS), MoS$_2$, a prototypical
transition metal dichalcogenide (TMD), shows a transition from an
indirect band gap of $1.3$ eV in a bulk structure to a direct band
gap of $1.8$ eV in the monolayer structure
~\cite{mak10,giacometti11}. The electronic structure of
ML-MDS exhibits a valley degree of freedom indicating that the
valence and conduction bands consist of two degenerate valleys
($K$, and $K'$) located at the corners of the hexagonal Brillouin zone. The lack of inversion symmetry of
ML-MDS results in a strong spin-valley coupling and the valence and conduction bands can be described by a minimal effective model Hamiltonian with a strong spin-orbit interaction which
splits the valence band into spin-up and spin-down subbands~\cite{xiao12,cui12,cao12}.
Due to the peculiar band structure, a variety of nanoelectronic
applications~\cite{wang12} including valleytronics, spintronics, optoelectronics and room
temperature transistors~\cite{giacometti11} have been
suggested for ML-MDS. Induction of valley polarization using
optical pumping with circularly polarized light is validated by
both {\it ab initio} calculations and experimental
observations~\cite{cui12,cao12,sallen12,mak12,behnia}. Also a combined
valley and spin Hall physics has been predicted as a result of
intimate coupling of the spin and valley degrees of
freedom~\cite{xiao12}.

In this work, we propose an effective model Hamiltonian
governing the low-energy band structure of monolayer TMDs
and show that its electronic properties
can be tuned by applying a perpendicular gate voltage.
Although our analysis here is focused on ML-MDS,
our approach can be easily generalized to other TMDs
. We obtain a seven-band model (for each spin component) in which four of them are contributed mainly
from sulfur (S) $p$ orbitals and the three remaining mostly originate
from molybdenum (Mo) $d$ hybrids. Our theory describes the
conduction and spin-split valence bands in accordance with early
theoretical studies~\cite{Bromely72, matthesiss}, and recent density functional
theory calculations~\cite{Kadantsev12,kang2013,ellis-all} and shows energy
corrections to the band structure by trigonal warping. The physics of naonribbon, defects, impurities and so on can be
studied by our lattice Hamiltonian. Intriguingly, our two-band
model Hamiltonian incorporates terms which invalidate the massive Dirac fermion picture of the low-energy behavior in ML-MDS.
When the system is subjected to a perpendicular magnetic field a Zeeman-like interaction for valleys breaks the
valley degeneracy of Landau levels in contrast to the finding in Ref.~\onlinecite{niu12}. Next,
we introduce the effect of a perpendicular gate voltage which
leads to shifts in the chemical potentials of three sublayers
consisting of one Mo and two S layers. We show that a perpendicular gate voltage leads to a splitting of high energy
bands originating from the $p$ orbitals of S atoms. One of our main
findings is the possibility of tailoring the band gap, effective masses and valley splitting of the valence and conduction bands by
varying the induced potentials in the three sublayers.
\par
The paper is organized as follows. In Sec.~II we introduce a
lattice model Hamiltonian and its low-energy two-band Hamiltonian that will be used in calculating the electronic
properties. In Sec.~III
we present our analytical and numerical results for the dispersion
relation of the ML-MDS in the presence of a magnetic field or a perpendicular gate voltage. Section~IV contains
a brief summary of our main results.

\section{Theory and Model}

ML-MDS consists of one layer of Mo atoms
surrounded by two layers of S atoms in such a way that each Mo atom is
coordinated by six S atoms in a trigonal prismatic geometry and
each S atom is coordinated by three Mo atoms. The symmetry space group
of ML-MDS is $D^1_{3h}$ which contains the discrete symmetries
$C_3$ (trigonal rotation), $\sigma_v$ (reflection by the $yz$
plane), $\sigma_h$ (reflection by the $xy$ plane) and any of their
products~\cite{xiao12}. In addition to the symmetry of the lattice, It is essential to consider the local atomic orbitals symmetries. The trigonal prismatic symmetry dictates that the $d$
and $p$ orbitals split into three and two groups, respectively:
$\{d_{z^2}\}$, $\{d_{x^2-y^2},d_{xy}\}$, $\{d_{xz},d_{yz}\}$ and
$\{p_x,p_y\}$, $\{p_z\}$. The reflection symmetry along the $z$
direction allows the coupling of Mo $d_{xz}, d_{yz}$ orbitals with
only the $p_z$ orbital of the S atom, whose contribution at the valence
band maximum (VBM) and the conduction band minimum (CBM) located at the symmetry points is negligible
according to first principle calculations.~\cite{Kadantsev12}
Therefore the conduction band minimum is mainly formed from Mo $d_{z^2}$ orbitals and the valence  band maximum
is constructed from the Mo $\{d_{x^2-y^2},d_{xy}\}$ orbitals with
mixing from S $\{p_x,p_y\}$~( Refs. [\onlinecite{Kadantsev12,kang2013}]) in both cases.

We thus can construct the tight-binding Hamiltonian for ML-MDS by
using symmetry adapted states and assuming nearest neighbor
hopping terms;
\begin{eqnarray}\label{tight}
{\hat H}_{TB}&=&\sum_{i\mu\nu}\{ \epsilon_{\mu\nu}^a a^\dagger_{i\mu} a_{i\nu}+\epsilon_{\mu\nu}^b b^\dagger_{i \mu} b_{i\nu}+\epsilon_{\mu\nu}^{b'} {b'}^\dagger_{i \mu} b'_{i\nu} \}\nonumber\\&+&\sum_{\langle ij\rangle,\mu\nu} t_{ij,\mu\nu} a^\dagger_{i \mu} (b_{i\nu}+ b'_{i\nu})+H.c.,
\end{eqnarray}
Here $a$ and $b (b')$ indicate the Mo and S atoms in the up (down) layer,
respectively. The indices $\mu$ and $\nu$ show the orbital degrees of
freedom labelled as $\{1,2,3\}\equiv\{d_{z^2},d_{x^2-y^2}+ i
d_{xy},d_{x^2-y^2}- i d_{xy}\}$ and $\{1',2'\}\equiv\{p_x+ i p_y,
p_x- i p_y\}$ for Mo and S atoms, subsequently. Therefore the
matrices $\epsilon^{a}$, $\epsilon^{b} (\epsilon^{b'})$, and
$t_{ij, \mu\nu}=<a,i,\mu|H|b,j,\nu>$ are responsible for the on-site energies of Mo
and S atoms, and hopping between different neighboring sites in
the space of different orbitals, respectively. We do need to take into account
the overlap integrals, ${\cal S}$, defined similar to the hopping terms
of the Hamiltonian with elements $s_{ij, \mu\nu}=<a,i,\mu|b,j,\nu>$.

Due to the trigonal rotational symmetry of the Hamiltonian, the
on-site energy matrices take the diagonal form
$\epsilon^{c}_{\mu\nu}=\epsilon^{c}_{\mu}+U^{c}$ for
$c:=\{a,b,b'\}$, in which $\epsilon^{c}_{\mu}$ shows the intrinsic
value of the on-site energies for the corresponding orbital state
and $U^{c}$ indicates the potential shift induced by the perpendicular gate voltage. Moreover, the symmetry properties of the lattice
lead to only three independent on-site energies
$A_1\equiv\epsilon_1^a$, $A_2\equiv\epsilon_2^a=\epsilon_3^a$, and
$B\equiv\epsilon_{1'}^{b(')}=\epsilon_{2'}^{b(')}$. Accordingly
the symmetries imposed by $\sigma_v, \sigma_h, C_3$ result in
constraints on the number of hopping integrals (three parameters)
 and overlap integrals (three parameters) (see Appendix A
for details). For the sake of definiteness, we choose
$t_{11}$, $t_{22}$, and $t_{21}$ as hopping integrals between the orbital pairs $(1,1')$, $(2,2')$, and $(2,1')$ along the
$\delta_{1\pm}$ directions, respectively
and corresponding forms for the overlap integral elements. A good approximation is
provided by the Slater-Koster method~\cite{slater54} in which all of the
hopping and overlap integrals are written as a linear combinations
of the hopping integrals $V_{pd\sigma}$, $V_{pd\pi}$ and overlap integrals $S_{pd\sigma}$, and $S_{pd\pi}$
where $V_{pd\sigma}=<{\bf R}',p,\sigma|H|{\bf R},d,\sigma>$
and $S_{pd\sigma}=<{\bf R}',p,\sigma|{\bf R},d,\sigma>$, for instance. To complete our
effective Hamiltonian, we need to add spin-orbit interaction
(SOI) in the model which causes spin-valley coupling in the
valence band. The large SOI in ML-MDS can be approximately
understood by intra atomic contribution $H_{\rm SO}=\xi(r){\bf S.
L}$. We only consider only the most important contribution of the Mo atoms
which gives rise to the spin-orbit coupling term $H^{Mo}_{\rm
SO}=\lambda\,{\rm diag}\{0,s,-s\}$ in the basis of states
$\{1,2,3\}$ where $\lambda$ is the
spin-orbit coupling and $s=\pm$. To study the band structure
properties provided by our tight-binding model, we find its
$k$-space form as $\sum_{{\bf k}s}\psi^\dagger_{{\bf k}s} ({\cal H}  -E{\cal S} ) \psi_{{\bf k}s}$
with $\psi_{{\bf k}s}=(a_{{\bf k}s,1} , a_{{\bf
k}s,2},a_{{\bf k}s,3} , b_{{\bf k}s,1'},b_{{\bf k}s,2'} , b'_{{\bf
k}s,1'},b'_{{\bf k}s,2'})^{\mathrm{\top}}$ in which $c_{{\bf
k}s,\mu}=\sum_{i} c_{is\mu}  \exp(i{\bf k}\cdot{\bf R}_i)$
($c:=\{a,b,b'\}$) are the annihilation operators of electrons with
momentum ${\bf k}$, spin $s$, and  orbital degree $\mu$. The
Hamiltonian density ${\cal H}$ and overlap ${\cal
S}$ are obtained as,\begin{eqnarray} {\cal
H}=\begin{pmatrix}{\hat H}_{a}&{\hat H}_{t}&{\hat H}_{t}\\{\hat H}_{t}^\dagger&{\hat H}_{b}&0\\{\hat H}_{t}^\dagger&0&{\hat H}_{b'}\end{pmatrix},~
{\cal
S}=\begin{pmatrix}1&{\hat S}&{\hat S}\\{\hat S}^\dagger&1&0\\{\hat S}^\dagger&0&1\end{pmatrix},
\end{eqnarray}
with the on-site energy Hamiltonian, ${\hat H}_{a}=U^a {\mathbbm
1}_3+{\rm diag}(A1,A_{+},A_{-})$, ${\hat H}_{b}=(B+U^{b}){\mathbbm 1}_2
$, ${\hat H}_{b'}=(B+U^{b'}){\mathbbm 1}_2 $, the hopping matrix,
\begin{eqnarray}
{\hat H}_t=\begin{pmatrix}t_{11}f({\bf k},\omega)&-e^{-i\omega}t_{11}f({\bf k},-\omega)\\
t_{21}f({\bf k},-\omega)&t_{22}f({\bf k},0)\\-t_{22}f({\bf k},0)&-e^{i\omega}t_{21}f({\bf k},\omega)\end{pmatrix},
\end{eqnarray}
and the overlap matrix ${\hat S}$ defined similarly to ${\hat H}_t$ but with the
$t_{\mu\nu}$'s replaced by $s_{\mu\nu}$'s. Here,
$A_\pm=A_2\pm\lambda s$ and $f({\bf k},\omega)=e^{i{\bf k}\cdot{\bm
\delta}_1}+e^{i({\bf k}\cdot{\bm \delta}_2+\omega)}+e^{i({\bf k}\cdot{\bm
\delta}_3-\omega)}$ is the structure factor with $\omega=2\pi/3$,
in-plane momentum ${\bf k}=(k_x,k_y)$, and ${\bm \delta}_i$
($i=1,2,3$) the in-plane components of the lattice vectors $\delta_{i\pm}$.

Generally, our tight-binding model leads to seven bands for each
spin component, however, in the absence of external bias i.e.
$U^{c} = 0$ the symmetry between top and bottom S sublayers reduces
the number of bands to five. Two of them correspond to the
conduction and valence bands, from which we calculate the
effective electron and hole masses, energy gap, and valence band
edge. Moreover, since the conduction band minimum mostly comes from $d$
orbitals~\cite{Kadantsev12}, we assume $10\%$
mixing with $p$ orbitals for the conduction band. This assumption is in good agreement with the result reported in Ref.~[\onlinecite{Cappelluti13}] (for more details on the effect of the mixing percentage see Appendix B). This provides us with
five equations for seven unknown parameters based on the
values obtained from {\it ab initio} calculations and experimental
measurements. Furthermore, it is reasonable~\cite{Bromely72} to consider $s_{\mu\nu}/t_{\mu\nu}=0.1eV^{-1}$ which
reduces the number of unknown parameters to five. We consider the energy gap
$\Delta=1.9$eV, spin-orbit coupling $\lambda=80$meV, effective
electron and hole masses $m_e=0.37m_0$ and $m_h=-0.44m_0$ ($m_0$ is
the free electron mass)~\cite{walle12}, and
$E_{VBM}=-5.73$eV~\cite{Yunguo12}. Eventually, all parameters can be fixed and we then obtain the on-site
energies $A_1=-1.45$eV, $A_2=-5.8$eV, $B=5.53$eV and hopping
integrals $e^{i\pi/6}t_{11}=0.82$eV, $e^{-i\pi/6}t_{21}=-1.0$eV,
and $e^{-i\pi/2}t_{22}=0.51$eV. With these parameters, our tight-binding theory is completed.

Now, we present an effective
low-energy two-band continuum Hamiltonian governing the conduction and
valence bands around the $K$ and $K'$ points, by exploiting the L\"{o}wdin
partitioning method~\cite{winkler}. We first change our
nonorthogonal basis ($|\psi\rangle$) to an orthogonal one
($|\psi^\prime\rangle={\cal S}^{1/2}|\psi\rangle$), leading to a
standard eigenvalue problem
$\widetilde{H}|\psi^\prime\rangle=E|\psi^\prime\rangle$ with
$\widetilde{H}={\cal S}^{-1/2}{\cal H} {\cal S}^{-1/2}$. More analytical calculations can be found in the Appendixes. To
employ the partitioning method, we expand the Hamiltonian
up to the second order in ${\bf q}={\bf k}-{\bf K}$ around
the ${\bf K}=(4\pi/3\sqrt{3}a_0,0)$ point which can be written as the sum of
$q$ independent (${\cal H}_0$) and $q$ dependent
(${\cal H}_1$) parts, ${\cal H}={\cal H}_0+{\cal H}_1$.
Then we rotate the orbital basis to
$|\psi'_{0i}\rangle$ ($i=1,..,5$) which are the eigenstates of
$\widetilde{H}_0$ corresponding to the eigenvalues $E_{i}$'s. In
the new basis, the transformed Hamiltonian is $U_0^\dagger  \widetilde{H} U_0$
where $U_0$ is the unitary diagonalizing matrix. We define two
subspaces $\{|\psi'_{01}\rangle,|\psi'_{02}\rangle\}$
corresponding to the conduction and valence bands and
$\{|\psi'_{03}\rangle,|\psi'_{04}\rangle,|\psi'_{05}\rangle\}$ for
the three remaining bands. Then we take the block diagonal and
off-diagonal parts of the Hamiltonian in these subspaces as
$H_{\rm diag}$ and $V$, respectively and use the unitary
transformation $H'=e^{-\cal O} U_0^\dagger  \widetilde{H} U_0 e^{\cal O}$ such that the lowest
order in $V$ is eliminated. This results in an effective
Hamiltonian $H'=H_{\rm diag}+[V,{\cal O}]/2$ which is block
diagonal in the subspaces up to the second order in $V$.
\par
The final result for the two-band Hamiltonian describing the
conduction and valence bands reads,
\begin{eqnarray}
H_{\tau s}&=& \frac{\Delta}{2}\sigma_z+\lambda\tau s
\frac{1-\sigma_z}{2}+t_0 a_0 {\bf q}\cdot{\bm \sigma}_\tau\nonumber\\
&+&\frac{\hbar^2|{\bf q}|^2}{4m_0}(\alpha+\beta\sigma_z)+t_1 a_0^2
{\bf q}\cdot{\bm \sigma}^{\ast}_{\tau}\sigma_x {\bf q}\cdot{\bm
\sigma}^{\ast}_{\tau}\end{eqnarray}
for spin $s=\pm$ and valley
$\tau=\pm$, with Pauli matrices ${\bm \sigma}_{\tau}=(\tau\sigma_x,\sigma_y)$ and momentum
${\bf q}=(q_x, q_y)$. The numeric values of the two-band model parameters are $t_0=1.68$eV, $t_1=0.1$eV, $\alpha=0.43$,
and $\beta=2.21$. Notice that $\alpha=m_0/m_{+}$ and $\beta=m_0/m_{-}-4m_0v^2/(\Delta-\lambda)$
where $m_{\pm}=m_e m_h/(m_h\pm m_e)$ and $v=t_0a_0/\hbar$.
Moreover, a quadratic correction $\delta\lambda\approx(0.03 eV) (a_0|{\bf q}|)^2$ arises to the spin orbit coupling due to folding down of the five-band model to a two-band one. This correction is estimated by using the effective masses of two spin-split valence band branches as $m_h(\uparrow)=-0.44m_0$ and $m_h(\downarrow)=-0.46m_0$ at the $K$ point. Notice that the correction term can be safely ignored in the validity range of the effective low-energy two-band model, ($a_0|{\bf q}|\ll 1$).
\begin{figure}
\includegraphics[width=1\linewidth]{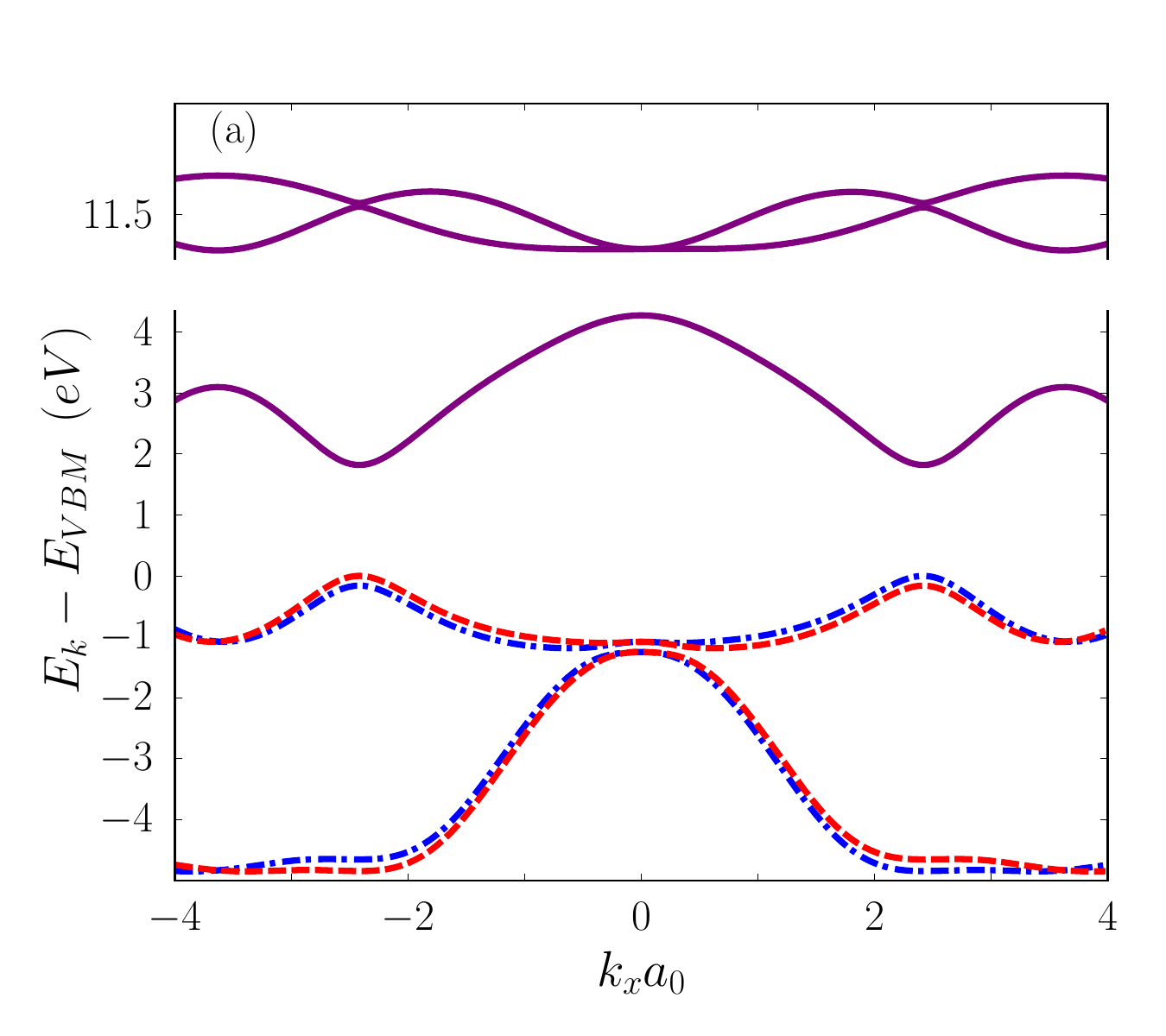}
\includegraphics[width=0.75\linewidth]{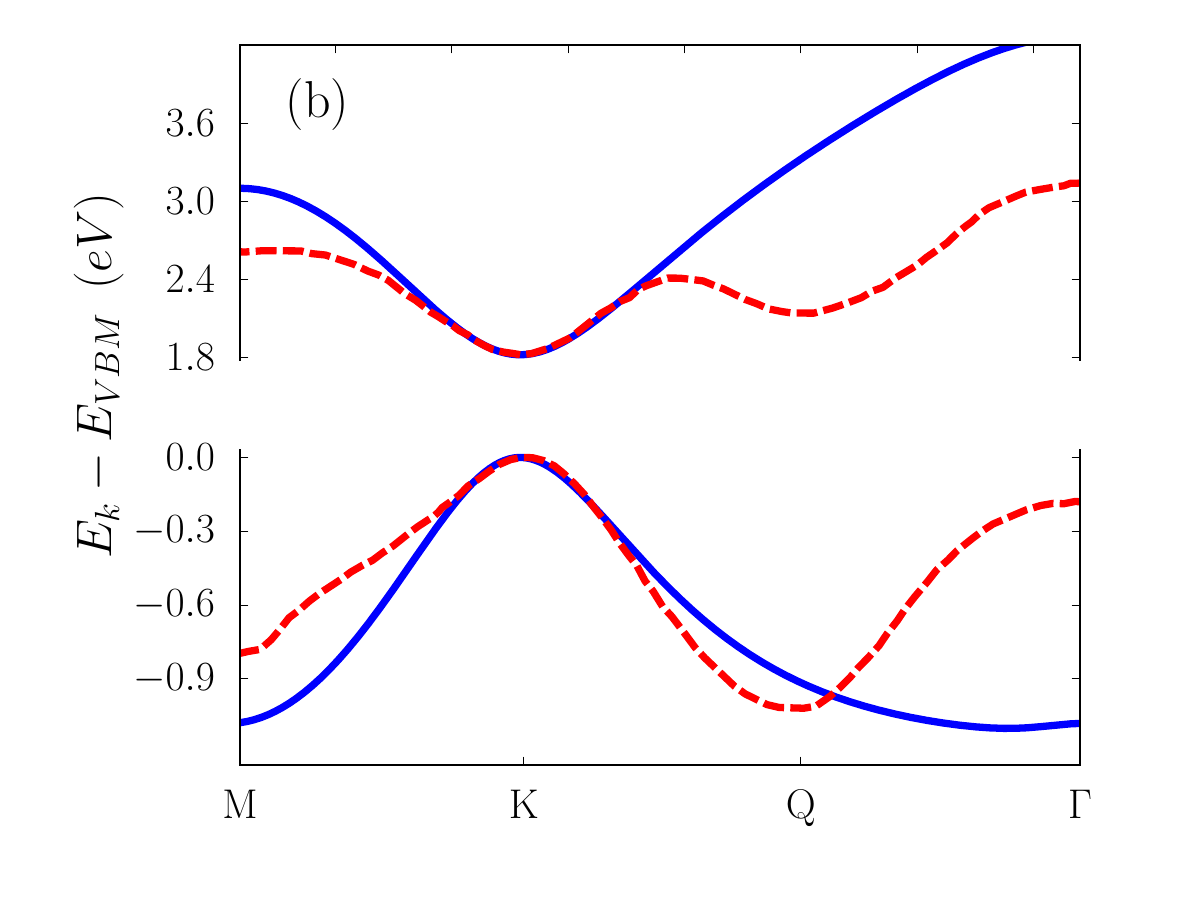}
\includegraphics[width=.22\linewidth]{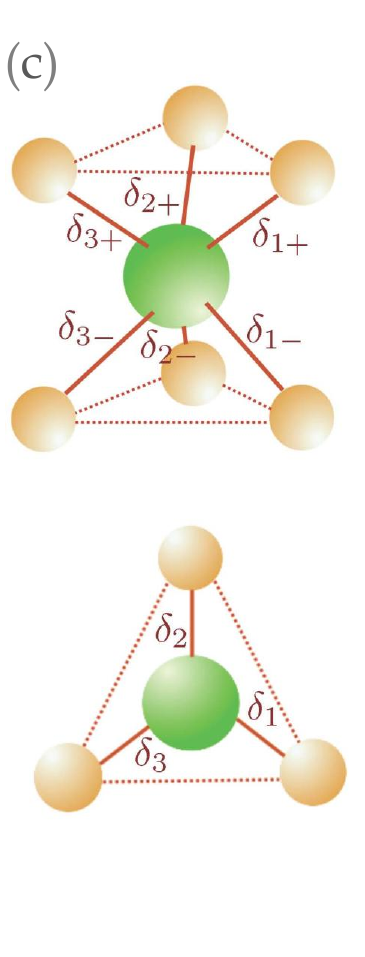}
\caption{(Color online) (a) Band structure of ML-MDS consisting of five bands in which
two are spin split in the valence band. The dot-dashed line refers
to one spin and the dashed line denotes another spin component. Solid lines refer to the spin degenerate band.
(b) A comparison between the band structure calculated by the present theory (solid lines) and the results calculated in Ref.~[\onlinecite{walle12}] (dashed lines) based on density functional theory. Notice that our theory works quite well around the ${\bm K}$ point for the particle (hole or electron) density less than $10^{14}$cm$^{-2}$ ($E_{F}-E_{CBM}\simeq 0.2$ eV). Here, $a_0=a \cos\theta$ where $a$ is the length of Mo-S bond and $\theta$ is the angle between the bond and the $xy$ plane. (c) Side and top views of the lattice structure are seen where Mo atom (larger green sphere) is surrounded by six S atoms.
}
\label{fig1}
\end{figure}
\par
The Hamiltonian differs from that introduced
by Xiao \emph{et al.}~\cite{xiao12} because of the second order terms
in $q$. The diagonal $q^2$ terms, which contribute in the energy,
to the same way as does the first order off-diagonal term, are
responsible for the difference between electron and hole masses
recently reported by using {\it ab initio} calculations~\cite{walle12}.
Moreover, the last term leads to anisotropic $q^3$ corrections
to the energy which contribute to the trigonal warping effect. Importantly, $\alpha$ vanishes for the case that $m_e=-m_h$,
however $\beta$ remains a constant.
Basically, there is the possibility to have a cubic off-diagonal term in the low-energy Hamiltonian which in the calculation
of the eigenvalues of the Hamiltonian are multiplied with the
off-diagonal $q$ terms and eventually contributes at the same
order as the diagonal $q^2$ terms. Since that term is very small, we thus ignore the $q^3$ off-diagonal term.

\section{Numerical results and Discussions}
In this section, we present our main calculations for the
electronic properties of MoS$_2$ by evaluating
Eqs.~(2), (3) and (4). We propose first the lattice
Hamiltonian by considering numerical values of the hopping integrals
and show the band structure of ML-MDS. Second, we present our numerical results for the electronic structure in two different models by exploring the Landau levels (LLs) and investigating the tunability of the electronic structure
via an external perpendicular gate voltage.

Figure~\ref{fig1} shows the band structure of ML-MDS consisting of
five bands for each spin in the absence of an external field. Two of them
are spin polarized (dot-dashed and dashed lines) and the others are spin
degenerate (solid lines). We note that due to the limitations of our
model, the high energy bands may not be comparable with those of
first principal calculations in a quantitative manner. Figure~\ref{fig1}(b) shows
a comparison between our results and those calculated by density functional theory~\cite{walle12} indicating that our theory is in good agreement with density functional theory results close to the ${\bm K}$ point up to a high particle (hole or electron) density $10^{14}$cm$^{-2}$ (the Fermi energy is $E_{F}-E_{\rm CBM}\simeq 0.2$ eV). Nevertheless our effective model Hamiltonian does not provide a good description of the physics around the $\Gamma$ point where other orbitals like $p_z$ must be considered in order to describe the electronic dispersion~\cite{Cappelluti13}.

\begin{figure}
\includegraphics[width=0.95\linewidth]{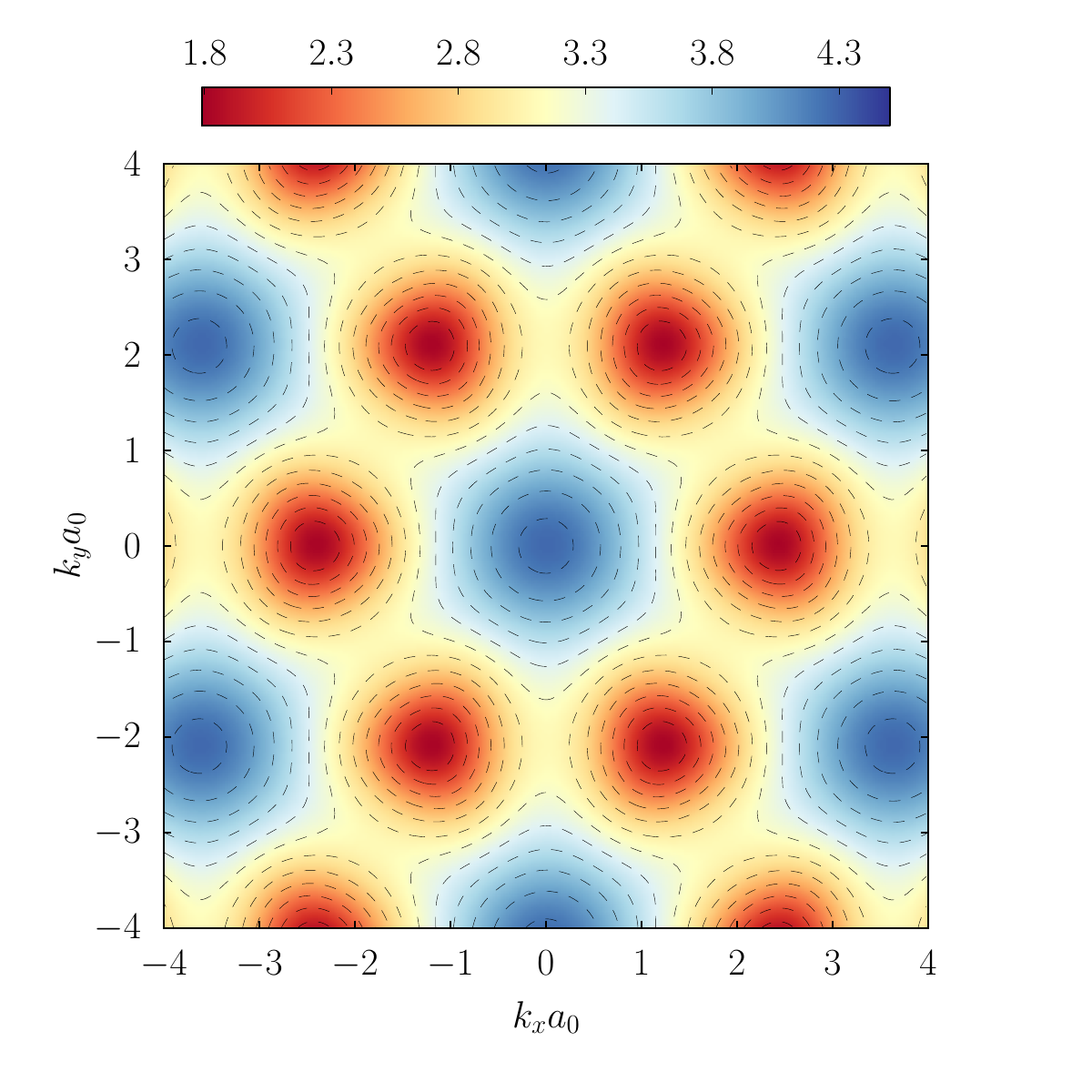}
\includegraphics[width=0.95\linewidth]{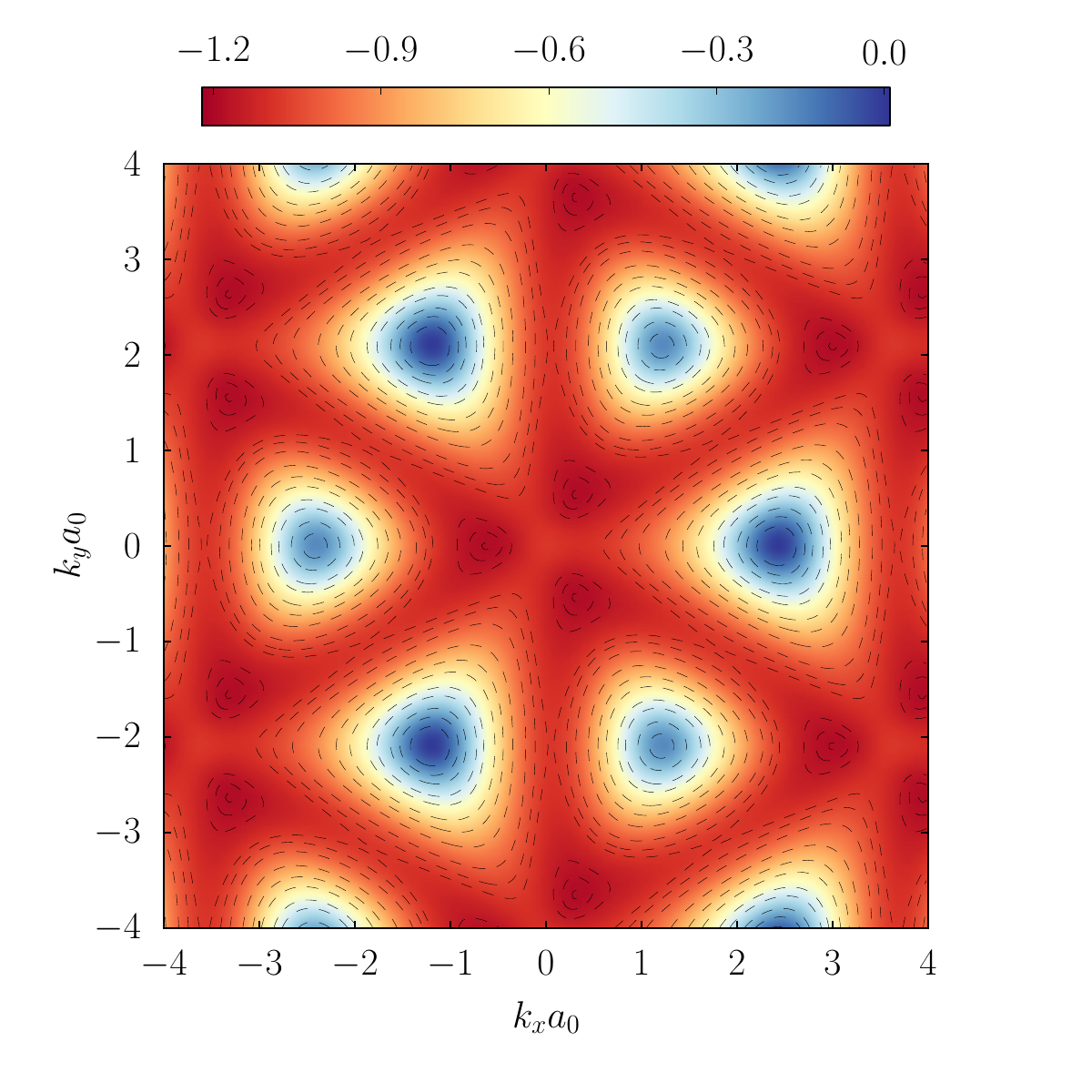}
\caption{(Color online) Contour plots of the conduction (top panel) and valence
(bottom panel) bands in momentum space for spin-up component together with isoenergy lines for to guide the eye.
While the conduction band shows almost isotropic dispersion, the trigonal warping occurs in
the valence band around $K$ points due to the difference of the orbital structures of the conduction and valance bands. }
\label{fig2}
\end{figure}

We further investigate the band structure close to the valence and conduction
bands and our numerical results are shown, via contour plots which show the isoenergy lines, in Fig.
\ref{fig2}. A strong anisotropy of the constant energy lines can be seen around
$K$ points in the valence band, due to the trigonal warping,
while in the conduction band all lines are almost isotropic; the warping is due to the difference of the orbital
structure of the conduction and valence bands.
\par
To study the interplay of spin and valley physics, we
introduce, by ignoring trigonal warping, the effect of a time reversal symmetry breaking term by
applying a perpendicular magnetic field, leading to the appearance
of LLs as follows,
\begin{eqnarray}
E^\pm_{n,\tau s}&=&\pm\sqrt{[\frac{\Delta-\lambda \tau s}{2}+ \hbar \omega_c (\beta n-\frac{ \alpha\tau }{2})]^2+2(\frac{t_0 a_0}{l_B})^2n}\nonumber\\
&&+ \frac{\lambda\tau s}{2}+\hbar \omega_c(\alpha n -\frac{\beta\tau}{2}),~~~n=0,1,\dots
\end{eqnarray}
where $\omega_c=eB/2m_0$ and $l_B=\sqrt{\hbar/(eB)}$ are the cyclotron
frequency and magnetic length, respectively.
It should be noticed that the trigonal warping term, $t_1$, leads to a second order perturbation correction in the Landau level energy and accordingly its effect on the Landau levels is very weak. In contrast to Ref.~\onlinecite{niu12}, we see an additional valley degeneracy breaking
term which is the reminiscent of the Zeeman-like coupling for
valleys. As a result, the
conduction band LLs are valley polarized and the valence
band LLs are both valley and spin polarized although we have not yet considered the
usual Zeeman interaction for spins. In particular, the $n=0$ LLs,
$E^+_{0,\tau s}=[\Delta -\hbar\omega_c \tau (\beta+\alpha)]/2$ and
$E^-_{0,\tau s}=\lambda\tau s-[\Delta +\hbar\omega_c \tau
(\beta-\alpha)]/2$, depend on the magnetic field strength in
opposite ways for the two valleys. More intriguingly, the splittings of
LLs in the conduction and valence bands $\delta
E^{+}\approx5.4\hbar\omega_c$ and $\delta E^{-}\approx4.6\hbar
\omega_c$, differ from each other due to the difference of $m_e$ and $-m_h$.
Furthermore, we can define extra splitting terms
for LLs in the conduction and valence bands: valley splitting
$\delta E^{+}_{v}=E^{+}_{-,s}-E^{+}_{+,s} = g_{v}^{+} \hbar
\omega_c$ and spin splitting $\delta
E^{+}_{s}=E^{+}_{\tau,-}-E^{+}_{\tau,+}= g_s \hbar \omega_c$ in the conduction band,
and
spin-valley splitting $\delta
E^{-}_{s-v}=E^{-}_{-,\bar{s}}-E^{-}_{+,s} = (sg_{s}+g_v^-) \hbar \omega_c$ in the valence band,  with $g_s=2$, $g_v^+\approx \beta+\alpha$ ($g_v^-\approx \beta-\alpha$) indicating spin and valley g-factors for the conduction and valence bands.
The splittings present in the conduction band,
originate from the valley and spin contributions, separately, but the splitting in the valence band comes from both spin and valley terms. Notice that the valley splitting depends slightly on the amount of  mixing with $p-$ orbitals for the conduction band through the parameter $\beta$ and the influence of the mixing value is very weak (see Appendix B for more details).

\begin{figure}
\includegraphics[width=0.9\linewidth]{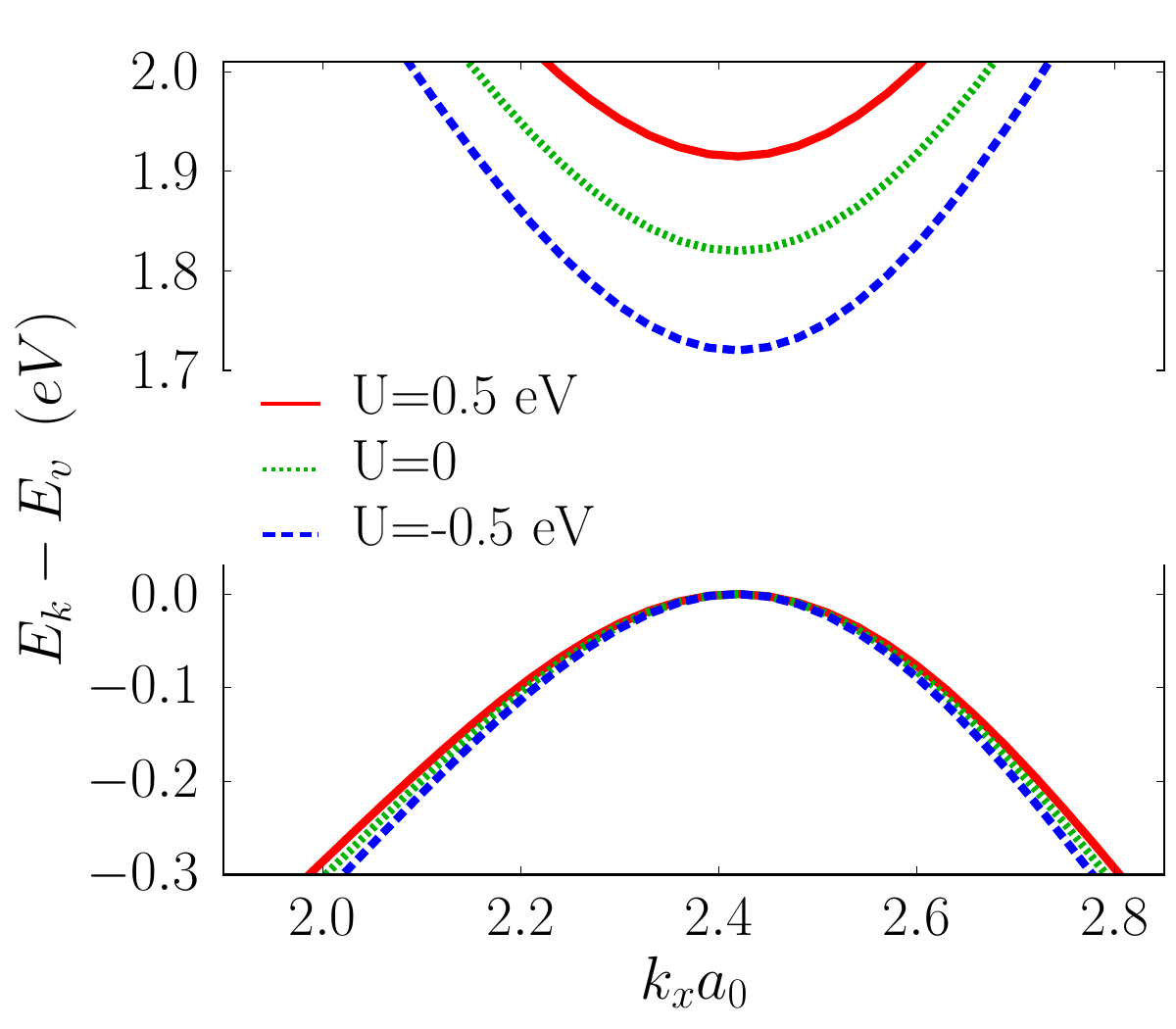}
\caption{(Color online) Spin-up conduction and valence
energy bands, referred to the energy $E_v=E_{VBM}+U/2$, for different values of the vertical bias.
} \label{fig3}
\end{figure}

\begin{figure}
\includegraphics[width=0.9\linewidth]{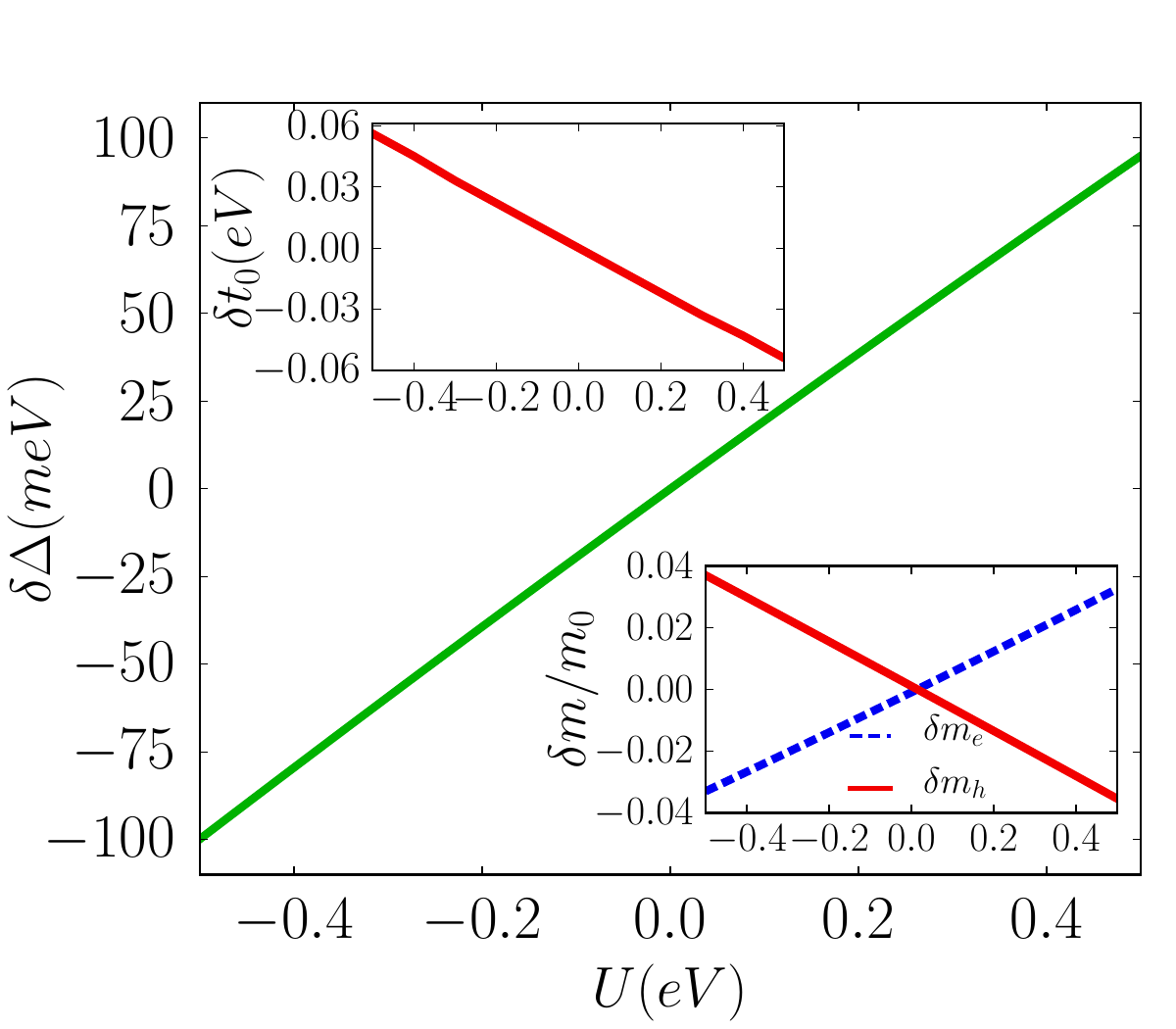}
\caption{(Color online) Variation of the gap versus induced potential
which is almost linear in the chosen range of $U$. Insets show the
change in the effective hopping integral and the effective masses with
$U$. Using the form of $U$-dependence of the parameters, we can also
estimate the linear change of $\delta \alpha\approx -0.15 U/$eV and
$\delta\beta\approx -1.95 U/$eV. } \label{fig4}
\end{figure}

It is also important to investigate the tunability of the electronic structure via a
perpendicular external electric field. The vertical bias breaks the
mirror symmetry $\sigma_h$ and modifies the on-site energies of
atoms in three sublayers of ML-MDS. Accordingly, these changes
affect the whole electronic structure especially the low-energy
characteristics such as $\Delta$, $m_e$, $m_h$ and
the effective hopping $t_0$. Interestingly enough, the valley degeneracy
breaking can be controlled by tuning $\alpha$ and
$\beta$ due to the perpendicular gate voltage when the system is subjected to a perpendicular magnetic field. We
assume a single-gated device in which the induced potentials take
the values $U^{b}=0$ and $U^{b'}=2U^{a}$. The variation of the mentioned
parameters with the induced potential $U^{b'}$ are shown in
Figs.~\ref{fig3} and \ref{fig4}, where we illustrate only the low-energy band structure for
different $U^{b'}$ values. Using simple electrostatic arguments,
the induced potentials for an applied vertical bias $V_g$ can be
estimated as $U^{b'}=\gamma eV_g$ with $\gamma\approx
(d/L)\epsilon'/\epsilon$ where  $\epsilon$, $d$ and $\epsilon'$, $L$ indicate the dielectric constants and thicknesses of ML-MDS and the substrate, respectively.
For typical values $\epsilon \approx6$~\cite{zhang02},
$\epsilon'\approx 3.9$, $d\approx 0.32$ nm, and $L\approx 300$ nm
with SiO$_2$ as the dielectric substrate, we obtain $\gamma\approx
0.0007$. By replacing the substrate with high-$\epsilon$ gate
dielectrics like HfO$_2$ with $\epsilon'\approx 25$
\cite{wallace}, the coefficient
$\gamma\approx0.004$ increases and leads to $U^{b'}\approx0.4$eV for
$V_g=100$V which is already used in the on-going experiments. Consequently,
the perpendicular gate voltage effects are enhanced by using a proper substrate with large dielectric constant.

\section{conclusion}

In summary, we have formulated a tight-binding Hamiltonian in order to
describe the low-energy band structure of monolayer MoS$_2$ which can
be useful to study energy dispersion and transport phenomena
in nanostructured MoS$_2$. We have obtained a seven-band model (for each spin component) in which four of them are contributed mainly
from sulfur (S) $p$ orbitals and three remaining mostly originate
from molybdenum (Mo) $d$ hybrids. Our model not only describes the low-energy behavior of monolayer MoS$_2$
which differs from massive Dirac fermion picture, but also predicts
the difference between the effective hole and electron masses and the
trigonal warping effect. In addition, the
two-band model leads to a
valley degeneracy breaking effect in the Landau levels and we have shown that the conduction band Landau levels are
valley polarized and the valence band Landau levels are both valley
and spin polarized. Finally, we have shown that by applying a
perpendicular electric field to the monolayer the electronic
structure especially the band gap and effective electron and hole
masses can be finely tuned. It should be noted that our model is appropriate mostly for low-energy
calculations in the vicinity of the conduction and valence bands.

Finally, we would like to emphasize that the diagonal quadratic terms in the low-energy Hamiltonian play an essential role in the transport and optical properties of the system~\cite{Rostami133}. It is worth  noting that the sign of the $\beta$ term can influence topological features of the system such as the Berry curvature, the valley Chern number which is defined as  ${\cal C}_v\propto {\rm sign}(\Delta)-{\rm sign}(\beta)$, and $Z_2$ invariant which vanishes for $\beta \Delta >0$.

{\it Note added--} Recently, a paper~\cite{Kormanyos13}
which covers closely related material, has been published.

\appendix

\section{Seven-band Hamiltonian}
We start constructing an effective tight-binding model for the monolayer MoS$_2$ system, assuming the following basis orbitals,
\begin{eqnarray}
|1\rangle&=&d_{z^2},\nonumber \\
|2\rangle&=&\frac{1}{\sqrt{2}}(|d_{x^2-y^2}\rangle+i |d_{xy}\rangle),
~~~|1'\rangle=\frac{1}{\sqrt{2}}(|p_x\rangle+i |p_y\rangle),\nonumber\\
|3\rangle&=&\frac{1}{\sqrt{2}}(|d_{x^2-y^2}\rangle-i
|d_{xy}\rangle),~~~|2'\rangle=\frac{1}{\sqrt{2}}(|p_x\rangle-i|p_y\rangle).\nonumber
\end{eqnarray}
The Wannier functions for different lattice sites in a crystal are
localized and they can be written as
$|c,i,\mu\rangle=c^\dagger_{i\mu}|0\rangle$ where $i$ denotes the
site and $\mu$ indicates atomic orbital. $c=a$ or $(b,b')$ shows
the annihilation operator of three different sublattices of
monolayer MoS$_2$, consisting of one Mo and two S atoms. Up to the
nearest-neighbor hopping integral, the tight binding Hamiltonian
can be written as Eq.~(1) in the main text,
\begin{eqnarray}\label{eq:eaebt}
\epsilon_{\mu\nu}^{c}&=&\langle c,i,\mu|H|c,i,\nu\rangle,\nonumber \\
t_{ij,\mu\nu}&=&\langle a,i,\mu|H|b(b'),j,\nu\rangle,  \\
s_{ij,\mu\nu}&=&\langle a,i,\mu|b(b'),j,\nu\rangle.\nonumber
\end{eqnarray}
\par
The lattice has two important symmetries $C_3=e^{-i\omega
L_z/\hbar}$ and $\sigma_v:\{(x,y,z)\rightarrow (-x,y,z)\}$ where
the first one is the trigonal rotational symmetry where
$\omega=2\pi/3$ and $L_z$ is the $z-$component of orbital angular
momentum and the second one indicates the reflection symmetry with
respect to the $y-z$ plane. The action of these symmetry operators on the
basis functions can be summarized in the following equations:
\begin{eqnarray}\label{eq:c3}
&C&_3|1\rangle=|1\rangle ~,~~~~~~~~~~~ C_{3}\{
|2\rangle,|2'\rangle\}=e^{i\omega}\{ |2\rangle,|2'\rangle\}\nonumber\\
&C&_{3}\{ |3\rangle,|1'\rangle\}=e^{-i\omega}\{
|3\rangle,|1'\rangle\}, \nonumber\\
&\sigma&_{v}\{ |1\rangle, |2\rangle,|3\rangle\}=\{|1\rangle,
|3\rangle,|2\rangle\}\nonumber\\
&\sigma&_{v}\{
|1'\rangle, |2'\rangle\}=-\{|2'\rangle, |1'\rangle \}.
\end{eqnarray}
It should be noticed that, here, we have dropped the spin indices.
The symmetry relations in Eq.~(\ref{eq:c3}), impose some
constraints on the on-site energies and hopping integrals, and thus we
have
\begin{eqnarray}\label{eq:ec3}
&&\epsilon^a=\begin{pmatrix}A_1&0&0\\0&A_2&0\\0&0&A_2\end{pmatrix}
\,,~~~~~~ \epsilon^b=\epsilon^{b'}=\begin{pmatrix}B&0\\0&B\end{pmatrix}\,,\nonumber\\
&&t_{\delta_{1\pm}}=\begin{pmatrix}t_{11}&-e^{-i\omega}t_{11}\\t_{21}&t_{22}\\-t_{22}&-e^{i\omega}
t_{21}\end{pmatrix},~~t_{\delta_{2\pm}}=\begin{pmatrix}e^{i\omega}t_{11}&-e^{i\omega}t_{11}\\e^{-i\omega}t_{21}&t_{22}\\-t_{22}&-e^{-i\omega}t_{21}\end{pmatrix}\nonumber\\
&&t_{\delta_{3\pm}}=\begin{pmatrix}e^{-i\omega}t_{11}&-t_{11}\\e^{i\omega}t_{21}&t_{22}\\-t_{22}&-t_{21}\end{pmatrix}.
\end{eqnarray}
Note that the same relations can be found for $s_{\delta_{i\pm}}$
by substituting the $t_{\mu\nu}$'s with $s_{\mu\nu}$'s. In the presence of
spin-orbit interaction of the Mo atoms, it is easy to generalize $\epsilon^a$ by replacing $(\epsilon^a)_{22}\rightarrow A_2 +
\lambda s$ and $(\epsilon^a)_{33}\rightarrow A_2 - \lambda s$. The subindices of
the hopping matrices indicate the nearest-neighbor vectors,
\begin{eqnarray}
\delta_{1\pm}&=&a(\frac{\sqrt{3}}{2}\cos\theta,-\frac{1}{2}\cos\theta,\pm\sin\theta),\nonumber \\
\delta_{2\pm}&=&a(0,\cos\theta,\pm\sin\theta),\\
\delta_{3\pm}&=&a(\frac{-\sqrt{3}}{2}\cos\theta,-\frac{1}{2}\cos\theta,\pm\sin\theta),\nonumber
\end{eqnarray}
where $a=2.43{\AA}$ and $\theta=40.7$ degree~\cite{Yue12} are
the Mo-S bond length and the angle between the bond and Mo's
plane, respectively. To find above equations, we also used
the operation of $C_3$ and $\sigma_v$ on $\delta_{i\pm}$  as
$C_3\delta_{1\pm}=\delta_{2\pm}$,
$C_3\delta_{2\pm}=\delta_{3\pm}$, $\sigma_v
\delta_{1\pm}=\delta_{3\pm}$ and $\sigma_v
\delta_{2\pm}=\delta_{2\pm}$.
\par
Following a method proposed by the Slater and Koster \cite{slater54}
(SK), all hopping and overlap integrals can be written as linear
combinations of $V_{pd\sigma}$ and $V_{pd\pi}$ and $S_{pd\sigma}$
and $S_{pd\pi}$. In this method, we define some standard hopping
and overlap parameters as $V_{ll'|m|}=\langle\vec R',l',m|H|\vec
R,l,m\rangle$ and $S_{ll'|m|}=\langle\vec R',l',m|\vec
R,l,m\rangle$ where $m=0,1$ stands for $\sigma,\pi$ bonds and the other
hopping and overlap integrals can be found from the SK
table~\cite{slater54}. In this way we find
\begin{eqnarray}
t_{11}&=&-\frac{1}{\sqrt{2}}(n_x + i n_y) [\sqrt{3} n_z^2 V_{pd\pi} + \frac{1}{2}(1-3n_z^2) V_{pd\sigma}],\nonumber \\
t_{21}&=&\frac{1}{2}(n_x-i n_y) [\frac{\sqrt{3}}{2} V_{pd\sigma} (1-n_z^2) + V_{pd\pi} (1+n_z^2)],\nonumber\\
t_{22}&=&\frac{1}{2}(n_x-i n_y)^3
(\frac{\sqrt{3}}{2}V_{pd\sigma}-V_{pd\pi}),
\end{eqnarray}
where ${\bf n}=(n_x,n_y,n_z)$ is a unit vector pointing
between the nearest-neighbor lattice points. Once again, we can
obtain similar relations for the overlaps by substituting the
hopping matrix elements with those of overlap matrix. These
relations help us to reduce the number of independent hopping
parameters from three to two. In the absence of external bias, due
to the symmetry between the two sulfur sublattices, we can simply
reduce the $7\times 7$ Hamiltonian to $5\times 5$ as below
\begin{eqnarray}
\cal{H}=\begin{pmatrix}{\hat H}_{a}&2{\hat H}_{t}\\2{{\hat H}_{t}}^\dagger&2{\hat H}_{b}\end{pmatrix},~~~~~
\cal{S}=\begin{pmatrix}1&2{\hat S}\\{2{\hat S}}^\dagger&2\end{pmatrix}.
\end{eqnarray}
\par
To find the unknown parameters, we should first obtain the energy
bands around the $K$-point. After solving the generalized eigenvalue
problem as $\cal H|\psi\rangle =\textit{E} \cal S|\psi\rangle$ at
the $K$-point, we can find the energies as
\begin{widetext}
\begin{eqnarray}\label{eq:E55}
E_1&=&A_{+},\nonumber\\
E_2&=&\frac{(A_1+B)/2-18\Re e(t_{11}s_{11}^*)-
\sqrt{[ (A_1+B)/2 -18\Re e(t_{11}s_{11}^*)]^2+(1-18|s_{11}|^2)(18|t_{11}|^2-A_1 B)}}{1-18|s_{11}|^2},\nonumber\\
E_3&=&\frac {(A_{-}+B)/2-18\Re e(t_{21}s_{21}^*)-
\sqrt{[(A_{-}+B)/2-18\Re e(t_{21}s_{21}^*)]^2+(1-18|s_{21}|^2)(18|t_{21}|^2-A_{-} B)}}{1-18|s_{21}|^2}, \\
E_4&=&\frac{(A_1+B)/2-18\Re e(t_{11}s_{11}^*)+
\sqrt{[ (A_1+B)/2-18\Re e(t_{11}s_{11}^*)]^2+(1-18|s_{11}|^2)(18|t_{11}|^2-A_1 B)}} {1-18|s_{11}|^2},\nonumber\\
E_5&=&\frac{(A_{-}+B)/2-18\Re e(t_{21}s_{21}^*)+
\sqrt{[(A_{-}+B)/2-18\Re e(t_{21}s_{21}^*)]^2+(1-18|s_{21}|^2)(18|t_{21}|^2-A_{-}
B)}}{1-18|s_{21}|^2}.\nonumber
\end{eqnarray}
\end{widetext}
These eigenvalues include bonding (with lower energy) and
anti-bounding ( with higher energy) of $p-d$ bands. Since the
conduction and valence bands are mostly formed from $d_{z^2}$ and
$d_{x^2-y^2}+i d_{xy}$ orbital of Mo, therefore, $E_1$ and $E_2$
are the valence band maximum and the conduction band minimum
located at the $K$-point, respectively.

\section{Two-band Hamiltonian}
Here, we find the low-energy two-band effective Hamiltonian with the
L\"{o}wdin partitioning method~\cite{winkler}. As described in
the text, we change the nonorthogonal basis to an orthogonal one and
then rotate them by using a unitary transformation, $U_0$, which
diagonalize $\widetilde{H}_0$ so that we arrive as a new
Hamiltonian,
\begin{eqnarray}
H=U^\dagger_0{\cal S}^{-1/2}{\cal H} {\cal S}^{-1/2} U_0,
\end{eqnarray}
where ${\cal H}={\cal H}_0+{\cal H}_1$ and up to the second order in $q=q_x+iq_y$, ${\cal H}_1$ is
given by
\begin{widetext}
\begin{eqnarray}
{\cal H}_1=\begin{pmatrix}
0&0&0&-\frac{3}{2}e^{i\omega}t_{11}|q|^2a^2_0&-3e^{i\omega}t_{11}q a_0\\
0&0&0&3e^{i\omega}t_{21}q a_0&-3t_{22}q^* a_0\\
0&0&0&3t_{22}q^* a_0&-\frac{3}{2}e^{-i\omega}t_{21}|q|^2a^2_0\\
-\frac{3}{2}e^{-i\omega}t_{11}^*|q|^2a^2_0&3e^{-i\omega}t_{21}^*q^* a_0&3{t_{22}}^*q a_0&0&0\\
-3e^{-i\omega}{t_{11}}^*q^* a_0&-3{t_{22}}^*q a_0&-\frac{3}{2}e^{i\omega} {t_{21}}^*|q|^2a^2_0&0&0
\end{pmatrix},
\end{eqnarray}
\end{widetext}
where $a_0=a\cos\theta$. In order to find the unitary transformation matrix, the eigenvectors
of $\widetilde{H}_0={S_0}^{-1/2} {\cal H}_0 {S_0}^{-1/2}$ are
obtained first. Fortunately, ${\cal S}_0^{-1/2}$ can be
analytically calculated at the $K$ point, however, for nonzero values
of $q$, an iterative method~\cite{Denman76} should be used.
Therefore, in the vicinity of  the $K$-point, we can treat ${\cal
H}_1$ as a perturbation by assuming small enough $q$ values and
the transformed Hamiltonian, $H$, can be expressed as the sum of
two parts where $V$ is a perturbation term,
\begin{eqnarray}
&&H_{\rm
diag}=\begin{pmatrix}h_{11}[2\times2]&0\\0&h_{22}[3\times3]\end{pmatrix},\nonumber\\
&&V=\begin{pmatrix} 0&h_{12}[2\times3]
\\{h_{12}}^\dagger[3\times2]&0\end{pmatrix}.
\end{eqnarray}
\par
Now, we can employ the quasi-degenerate perturbation theory that
is based on the idea of constructing a unitary operator $e^{-\cal
O}$ in such a way to drop the first-order $V$ in the transformed
Hamiltonian, $H'=e^{-\cal O} H {e^{\cal O}}=H_{\rm diag}+V+[H_{\rm
diag},{\cal O}]+[V ,{\cal O}]+\frac{1}{2}[[H_{\rm diag}, {\cal
O}], {\cal O}]+\cdots$. This imposes the constraint
$V+[H_{diag},{\cal O}]=0$ which leads to the following form for
the generator of the transformation,
\begin{eqnarray}
{\cal
O}=\begin{pmatrix}0&\eta[2\times3]\\-\eta^\dagger[3\times2]&0\end{pmatrix},
\end{eqnarray}
with the property that $\eta h_{22}- h_{11}\eta=h_{12}$. Then
$ H'= H_{\rm diag} +\frac{1}{2}[V,{\cal O}]+\cdots$ is an
effective Hamiltonian with two decoupled subspaces. Following
straightforward calculations, the effective Hamiltonian of the
low-energy bands can be obtained as follows
\begin{eqnarray}
h_{11}-\frac{1}{2}\{\eta {h_{12}}^\dagger+h_{12}\eta^\dagger\}.
\end{eqnarray}
Insertion of matrix form of operators $h_{11}$, $h_{12}$, and
$\eta$ results in the two-band form proposed in Eq.~(4) in the
text as
\begin{eqnarray}
H_{\tau s}&=& \frac{\Delta}{2}\sigma_z+\lambda\tau s
\frac{1-\sigma_z}{2}+t_0 a_0 {\bf q}\cdot{\bm \sigma}_\tau\nonumber\\
&+&\frac{\hbar^2|{\bf q}|^2}{4m_0}(\alpha+\beta\sigma_z)+t_1 a_0^2
{\bf q}\cdot{\bm \sigma}^{\ast}_{\tau}\sigma_x {\bf q}\cdot{\bm
\sigma}^{\ast}_{\tau}\end{eqnarray} for spin $s=\pm$ and valley
$\tau=\pm$.

It should be noticed that the parameters of the tight-binding model dependence on the $p-$orbital mixing do not have a simple form which prevents us from introducing an analytical relation between the parameters of the two-band model Hamiltonian and the $p-$orbital percentage. The parameter $\alpha$ depends only on the effective mass difference between the conduction and valence bands. The energy gap and the spin orbit splitting do not depend on the mixing. Therefore, the influence of the mixing parameter has been checked for $5\% $ and $15\%$ mixing and they lead to the following values $\beta= 2.30$ and $2.15$, $t_0=1.65$ and $1.70$, respectively. In addition, the values of the effective hopping trigonal wrapping $t_1$ in the low-energy Hamiltonian change slightly with variation in the mixing value by $1\%$ which can be neglected.

\end{document}